\documentclass[twocolumn,aps,showpacs,prl,superscriptaddress]{revtex4}%,showpacs]{revtex4}
\usepackage{graphicx}
\usepackage{dcolumn}% Align table columns on decimal point
\usepackage{bm}% bold math
 \usepackage{amssymb}

\begin{document}

\title{A Maximum-Likelihood Analysis of Observational Data on Fluxes and Distances of Radio Pulsars:
Evidence for Violation of the Inverse-Square Law}  

\author{John Singleton}%\email{jsingle@lanl.gov}
\affiliation{National High Magnetic Field Laboratory, Los Alamos National Laboratory, MS-E536, Los Alamos, NM 87545, USA}
\author{Pinaki Sengupta}
\affiliation{School of Physical and Mathematical Sciences, Nanyang Technological University, 
50 Naynag Avenue, Singapore 639798}
\author{John Middleditch}
\affiliation{CCS-3, MS B265, Los Alamos National Laboratory, Los Alamos, NM 87545, U.S.A.}
\author{Todd~L.~Graves}
\affiliation{CCS-6, MS F600, Los Alamos National Laboratory, Los Alamos, NM 87545, U.S.A.}
\author{Mario~R.~Perez}
\affiliation{Astrophysics Division, 3Y28, NASA-Headquarters, 300 E. Street SW, Washington DC 20546, U.S.A.}
\author{Houshang Ardavan}
\affiliation{Institute of Astronomy, University of Cambridge, Madingley Road,
Cambridge CB3 OHA, U.K.}
\author{Arzhang Ardavan}
\affiliation{Clarendon Laboratory, Department of Physics, University of Oxford, Parks Road,
Oxford OX1 3PU, U.K.}

\begin{abstract}
We analyze pulsar fluxes at 1400~MHz ($S_{1400}$) 
and distances ($d$) extracted from the
Parkes Multibeam Survey.
Under the assumption that distribution of pulsar luminosities is distance-independent, 
we find that either (a)~pulsar
fluxes diminish with distance according to a non-standard power
law, due, we suggest, to the presence of a component with $S_{1400} \propto 1/d$,
or (b)~that there are very significant ({\it i.e.} order of magnitude) 
errors in the dispersion-measure method for estimating pulsar distances.
The former conclusion (a) supports
a model for pulsar emission that has also successfully
explained the frequency spectrum of the Crab and 8 other pulsars over
16 orders of magnitude of frequency, whilst alternative (b) would necessitate
a radical re-evaluation of both the dispersion-measure method
and current ideas about the distribution of free electrons within our Galaxy. 
\end{abstract}


\maketitle
Recently, superluminal polarization currents, whose distribution patterns
move faster than light {\it in vacuo},
have been invoked as sources of pulsar emission~\cite{mnras}.
This idea is derived from the work of Bolotovskii, Ginzburg and others,
who showed both that such superluminal polarization currents do not violate
Special Relativity (since the oppositely-charged particles that make them move 
subluminally) and that they form a {\it bona-fide} source term in 
Maxwell's equations~\cite{bolot1,ginz,CandL,bolot2}.
The validity of these ideas has been demonstrated in a variety of laboratory
experiments~\cite{bess1,bess2,singleton,bolot3,ardaexp}. 
Moreover, by extending the approach to a superluminal polarization current 
whose distribution pattern 
follows a circular orbit,  it was possible to explain {\it quantitatively} 
several observables from the Crab pulsar, including the spacing and widths of 
the emission bands at frequencies around 8~GHz, 
the maximum of the radiation spectrum, and the 
overall continuum spectrum across 16 orders of magnitude in 
frequency~\cite{mnras}. Subsequently, successful
quantitative fits were carried out for 8 other pulsars~\cite{mnras2} and
a related superluminal model 
reproduced the general form of pulsar
Stokes parameters~\cite{schmidt}.

In this Letter, we demonstrate a further prediction for rotating superluminal sources; 
that there is a component of the emission
whose flux $S$ decays with distance $d$ 
as $S\propto 1/d$~\cite{morph,josa2004}, rather than the conventional inverse
square law ($S \propto 1/d^2$). 
Our demonstration employs a
Maximum Likelihood Method (MLM)~\cite{george} analysis of
pulsar observations~\cite{manchester}.
The MLM is carried out to circumvent the significant
Malmquist bias~\cite{malmquist} due to the increasing non-detection of
weaker pulsars as $d$ increases.

The pulsar emission component with $S \propto 1/d$ is due to
a general property of sources that travel faster
than their emitted waves; the 
relationship between reception time and 
retarded time is not monotonic and one-to-one~\cite{bolot2,lowson}. 
Multiple retarded times~\cite{gold2}, or, if the source accelerates, 
extended periods of retarded time~\cite{gold2}, 
can contribute to the waves received instantaneously,
resulting in {\it temporal focusing},
{\it i.e.}, concentration of the energy carried by the waves 
in the time domain~\cite{bolot3,ardaexp,morph}.
This effect is well known in acoustics~\cite{lowson,gold2,lowson2}.
It is the temporal focusing from the parts of the source that
approach the observer at the wave speed and with zero acceleration
that leads to the $S\propto 1/d$ flux component~\cite{morph,josa2004}.
Note that this mechanism does not violate
conservation of energy since the enhanced flux detected in
some places is compensated exactly by diluted fluxes 
elsewhere~\cite{morph,lowson}.
Moreover, we emphasize that
the emission discussed in this paper arises from true superluminal 
motion; electromagnetic disturbances (polarization currents)
that travel faster than the speed of light {\it in vacuo}, $c$~\cite{bolot3}. 
This should not be confused with the {\em apparent} 
superluminal motion of certain radio sources that
is thought to arise from relativistic aberration~\cite{ZandP}.

\begin{sloppypar}
Our analysis of $S_{1400}$ versus $d$
assumes that the luminosity function of pulsars is uniform
throughout the Galaxy, {\it i.e.}
that the populations of pulsars
at various distances from the Earth
are similar, each containing
a representative spread of pulsar types,
sizes and energies. 
We extract 1400~MHz fluxes ($S_{1400}$) and dispersion measures
from the ATNF
Pulsar Catalogue~\cite{manchester} 
(http://www.atnf.csiro.au/research/pulsar/psrcat). 
To eliminate statistical biases from
different instruments, we restrict the sample to
the 1109 galactic pulsars detected 
using a single instrument, the 
Parkes Multi-beam Survey~\cite{parkes}.  
We use the so-called NE2001~\cite{dm3} model
to evaluate $d$ values from the dispersion measures given in the ATNF catalogue; 
this was shown~\cite{dm3,giguere} to give
pulsar positions that are more consistent with the known
distributions of matter in the Galaxy than previous 
models.
\end{sloppypar}  
\begin{figure}
\centering
\includegraphics[width=1.1\columnwidth]{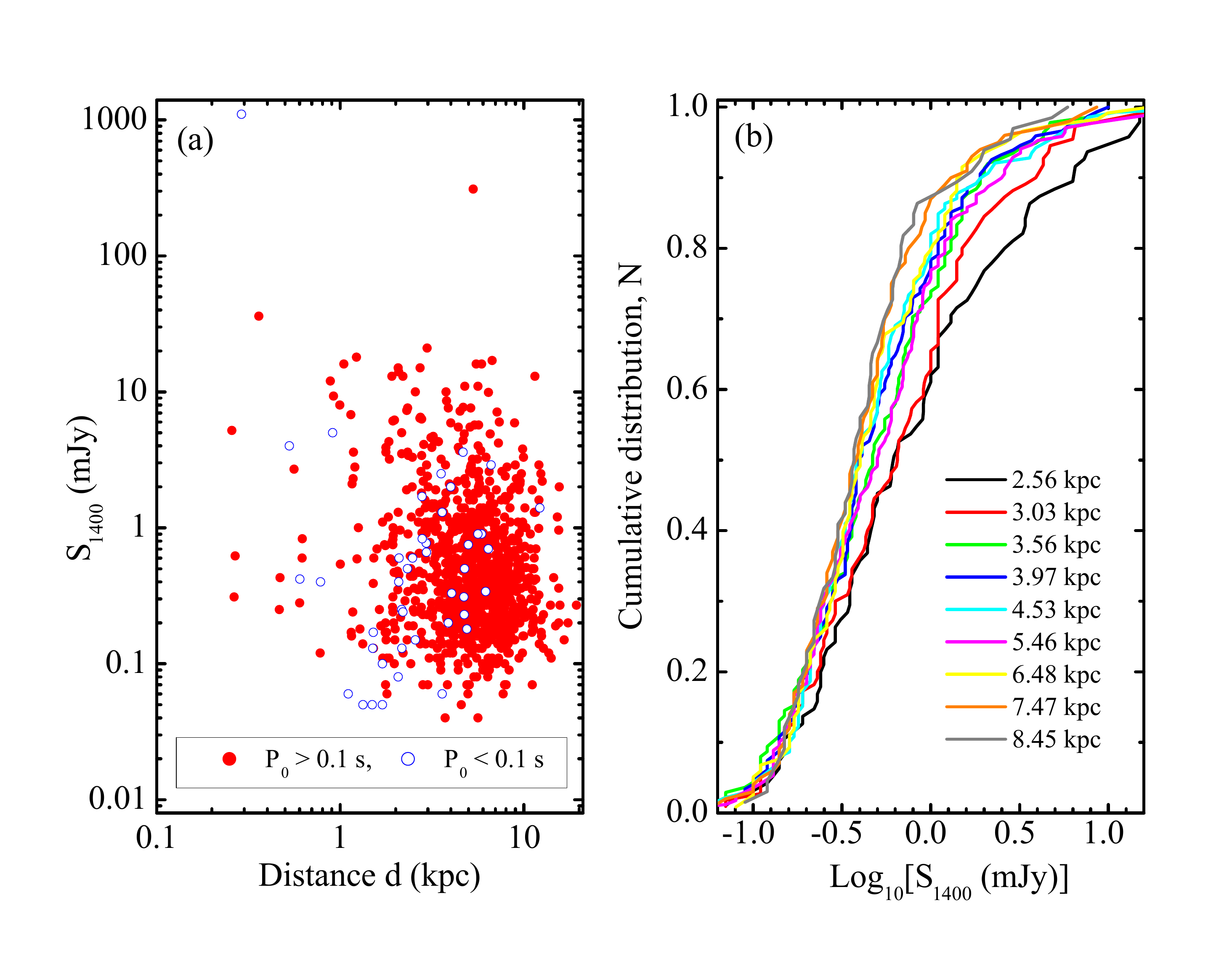}
\vspace{-10mm}
\caption{(a)~The 1109 Galactic pulsars in the Parkes Multibeam Survey
plotted as $S_{1400}$ versus
distance $d$, where $d$ is determined from the NE2001 interpretation
of dispersion measure~\cite{dm3}.
Pulsars with periods $P_0 < 0.1$~s are shown
as hollow points, and those with $P_0>0.1$~s
are displayed as filled points. 
The apparent differences between the distributions
of the two sets of pulsars may reflect the fall-off in sensitivity
of the Parkes instrument for faster pulsars (see 
Fig.~2 of Ref.~\cite{manchester2}).
(b)~Cumulative population distribution in $S_{1400}$ 
for 9 distance bins. 
The mean distance of each bin
is given in the inset key. } 
\vspace{-5mm}
\label{astFig1}
\end{figure}

We first show that the Parkes observations show
a strong Malmquist bias due to instrumental issues; 
consequently, the MLM~\cite{george} 
is essential in making quantitative conclusions
about the flux-distance relationship.
Fig.~1(a) plots the 
Galactic pulsars from the Parkes Multibeam Survey  as
$\log_{10}(S_{1400})$ versus $\log_{10}(d)$.
It is already obvious that data are very sparse for
$S_{1400} \leq 0.1$~mJy. To assess whether
this is an instrumental artefact, or
a fundamental property of the pulsar population,
we group the pulsar data in bins covering
certain distance ranges ({\it e.g.} $6.0\leq d\leq 7.0$~kpc)
and plot the cumulative distribution functions $N(S_{1400})$
of each bin in Fig.~1(b).
The $d$ bins are chosen so that they 
cover a reasonably small range of $d$ but
contain a large enough population for meaningful
statistics ($\sim100$ pulsars).

Note first that all the cumulative distribution functions in Fig.~1(b)
tend to zero at roughly the same $S_{1400}$. 
This strongly suggests that low-flux
part of each cumulative distribution is
representative of the roll-off in sensitivity
of the instrument, rather than an
intrinsic property of each pulsar population.
On the other hand, the high-flux sides of the curves in Fig.~1(b)
are likely to be more representative of
intrinsic properties of the pulsar populations.
As such, they should move to lower fluxes as $d$ increases.
This does indeed happen, but 
at a slower rate than the inverse-square law; the
75\% points of the functions spread over 
roughly a factor 2.8 in $S_{1400}$, even though
the distance varies by a factor of around 3.3.
This is a much smaller spread than that expected for the
inverse square law ($\sim 10$).

Simple analysis thus far has suggested
that the Parkes Multibeam Survey 
is subject to a substantial Malmquist bias
because it misses a large
fraction of pulsars with $S_{1400} \leq 0.4$~mJy
and cuts off completely for $S_{1400} < \sim 0.1$~mJy.
Both of the latter figures are of the same order of magnitude as the
calculated minimum detectable flux of the
Parkes instrument ($\sim 0.15$~mJy;
see Fig.~2 of Ref.~\cite{manchester2}).
To make further progress, we require a
method that attempts to compensate for missing data, caused by
instrumental sensitivity problems, in a systematic way.
Originally, Efstathiou {\it et al.}~\cite{george}, and subsequently a
number of authors,
have demonstrated that the MLM
is very suitable for such problems by applying it to 
red-shifts of very distant objects, a data set which is
incomplete due to instrumental problems somewhat analogous 
to those of the Parkes survey~\cite{george}.
The Parkes database is especially
suitable for treatment, since $S_{1400}$ and $d$ values 
are essentially
independently derived.

Our implementation of the MLM determines
the probable luminosity function, $\phi(L)$, based 
on the (incomplete) observed data, where $L$
is the luminosity.
The technique fits a quasi-continuous $\phi(L)$ to the observations 
under the assumption of an instrumental cut-off~\cite{george}. 
Here, the MLM
is used to determine the most likely
value of the exponent $n$ in the relationship
$S_{1400}\propto d^{-n}$.
The intrinsic luminosity function is therefore calculated
using the $S_{1400}$ and $d$ values from
Parkes Survey for each of the trial exponents 
($n_{\rm trial} \in \{0.5, 1.0, 1.5, 2.0, 2.5, 3.0\}$);
this feeds into the self-consistent determination of
$\phi(L)$ and the cutoff through the relationship
$L_i=S_id_i^n$, where  $S_i=S_{1400}$ for the $i^{\rm th}$ pulsar
at distance $d_i$.
The probability that pulsar $i$ 
is observed in a flux-limited survey is given by
\[
p_i \propto \phi(L_i)/\int_{L_{\rm min}(d_i)}^{\infty}\phi(L){\rm d}L.
\]
where $L_{\rm min}(d_i)$ is the minimum luminosity that
a pulsar at distance $d_i$
can have to be detected.
We define a likelihood function
${\cal L}= \prod_ip_i$.
Following Ref.~\cite{george},
we use an approach that does not assume a simple functional for
$\phi(L)$. Instead, the luminosity function is 
parameterized as $N_b$ steps:
$\phi(L)=\phi_k$, for $L_k-{\Delta L\over 2}<L<L_k+{\Delta L\over 2}$, with $k=1,...,N_b$.
The maximum likelihood function assumes the form
\[
\ln {\cal L} = \sum_{i=1}^{N_p}W(L_i-L_k)\ln \phi_k - 
\]
\begin{equation}
\sum_{i=1}^{N_p}\ln \left\{ \sum_{j=1}^{N_b}\phi_j\Delta LH[L_j-L_{min}(d_i)]\right\} + {\rm const}.
\end{equation}
where $N_p$ is the total number of pulsars in the Parkes survey.
Here, $W(x)=0$ for $ -\Delta L/2\leq x \leq \Delta L/2$, and 1 otherwise, and
\[
H(x)=\left\{\begin{array}{cl}
        0, & x \leq -\Delta L/2 \\
	(x/\Delta L + 1/2), & -\Delta L/2\leq x \leq \Delta L/2 \\
	1, & x \geq \Delta L/2.
	   \end{array}\right.
\]
The parameters $\phi_k$ determining the luminosity 
function are given by the self-consistent
set of equations
\[
\phi_k\Delta L = \frac{\sum_iW(L_i-L_k)}{\sum_i \frac{H[L_k-L_{\rm min}(d_i)]}{\sum_{j=1}^{N_b}\phi_j\Delta LH[L_j-L_{\rm min}(d_i)]}},
\]
with $k = 1,......,N_b$.
The above equations are solved iteratively to obtain the luminosity function,
with the goodness of fit being parameterized by the relative convergence error 
$\epsilon = \sum_{b=1}^{N_b}(\phi_i(b)-\phi_{i-1}(b))^2$.
This is basically the relative mismatch between successive iterations;
the smaller the value of $\epsilon$, the better the representation of the data.
In all cases, the $10-15$\% or so of pulsars with 
very high intrinsic luminosity 
were excluded from the 
analysis to ensure a quasi-continuous 
distribution of the luminosity function.

When the complete Parkes data set is used
(Fig.~2(a), solid points), we find that the derived luminosity 
function converges very rapidly for
$n_{\rm trial}$ = 1.0 and 1.5 with a small $\epsilon$. 
The convergence to a putative luminosity function 
is considerably ($\sim 10^5$) worse when one assumes an unphysical 
$n_{\rm trial}$ = 0.5, 2.5 and 3.0, as well as the commonly accepted
inverse-square law ($n_{\rm trial} =2$). 
The relative convergence for these values of $n_{\rm trial}$
can be somewhat improved by 
restricting the analysis to a smaller set of pulsars, but is still 
not comparable to those obtained for $n_{\rm trial} = 1.0$ and 1.5. 
Overall, the best combination is $n_{\rm trial} = 1$ 
with 983 pulsars fitted, implying
$S_{1400}\propto 1/d$; the error with $n_{\rm trial} = 1.5$
is somewhat larger, with 980 pulsars fitted.
However, the good convergence for both exponents
suggests that the observed flux may be a mixture
of the $S_{1400}\propto 1/d$ component and
spherically-decaying radiation, both of which are
to be expected from a superluminally-rotating source~\cite{morph}.
In slower pulsars, whose light cylinders lie further away
from the central neutron star, the superluminally-rotating
part of the current distribution may not be dense enough
to give rise to a dominant nonspherically-decaying component
of the radiation~\cite{mnras}.

The main assumption of our analysis thus far is that
pulsar populations are similar throughout the Galaxy.
This is potentially open to question
if there are two distinct populations of pulsars,
especially if some property of each population results
in different instrumental cut-offs. 
As shown in Fig.~2 of Ref.~\cite{manchester2},
the sensitivity of the Parkes Instrument is limited for
pulsars with periods $P_0 < 0.1$~s;  {\it i.e.,} distant
millisecond pulsars are harder to detect than equivalent 
slower pulsars (see Fig.~1(a)). This
might result in an apparent
$d$-dependent evolution of the characteristics of the detected pulsar population.
Second, though recent work~\cite{mnras,mnras2} suggests
that {\it all} pulsars possess the same emission mechanism,
some opine that millisecond pulsars form a distinct
population~\cite{LandGS} and might therefore possess a different luminosity function.
Both of these concerns can be addressed by excluding pulsars with 
periods $P_0<0.1$~ms from the MLM fit.
The hollow points in Fig.~2(a) show the result; 
the minimum $\epsilon$ is obtained with $n_{\rm trial}=1.5$.
For comparison, Fig.~2(b)
shows the result of running the MLM
on only the 43 pulsars with $P_0<0.1$~s; once again, the fit
is best for $n_{\rm trial} =1.5$.
Therefore, in spite of the separation of pulsars
with $P_0<0.1$~s and $P_0> 0.1$~s,
the MLM {\it never} obtains
a minimum $\epsilon$ for $n_{\rm trial} =2$;
the errors for $n_{\rm trial} = 1, 1.5$
are smaller, often dramatically so.
This suggests that the violation of the inverse-square law
by the pulsar population is a robust phenomenon.

Finally, the MLM was tested on 
two synthetic galaxies of pulsars
in which $S_{1400}$ was constrained to decay as $d^{-1.5}$
and $d^{-2}$ (Fig.~2(c)). 
As all pulsar population models in the literature ({\it e.g.} Refs.~\cite{arzoumanian,giguere})
are both contaminated with assumptions involving the
inverse-square law {\it and} involve $\sim 10-20$ adjustable
parameters, we derived
synthetic pulsar distributions that are consistent with
the Parkes database using Bayesian methods~\cite{todd}.
As with the Parkes data,
the luminosity function was determined
using a range of trial exponents ($n_{\rm trial} =0.5$, 1, 1.5, 2 and 2.5).
In each case, the MLM
located the correct value of $n$, with an accuracy better than 
$\pm 0.5$ (Fig.~2(c)). This
gives confidence in
our assertion that pulsar flux data in the Parkes survey
violate the inverse square law.

In the various implementations
of the MLM,
the inferred pulsar luminosity function ({\it e.g.}, Fig.~2(d)) always
decreases monotonically~\cite{econ}. As discussed above,
the effects of instrumental insensitivity will lead to an {\it apparent}
luminosity function that falls off at low $S_{1400}$.
This problem may have led to the commonly-held belief that
the pulsar luminosity function is a log-normal distribution~\cite{LandGS}.
Once one compensates for the loss of data (Fig.~2(d)),
the maximum
characteristic of a log-normal distribution will be absent.

\begin{figure}
\centering
\includegraphics[width=0.99\columnwidth]{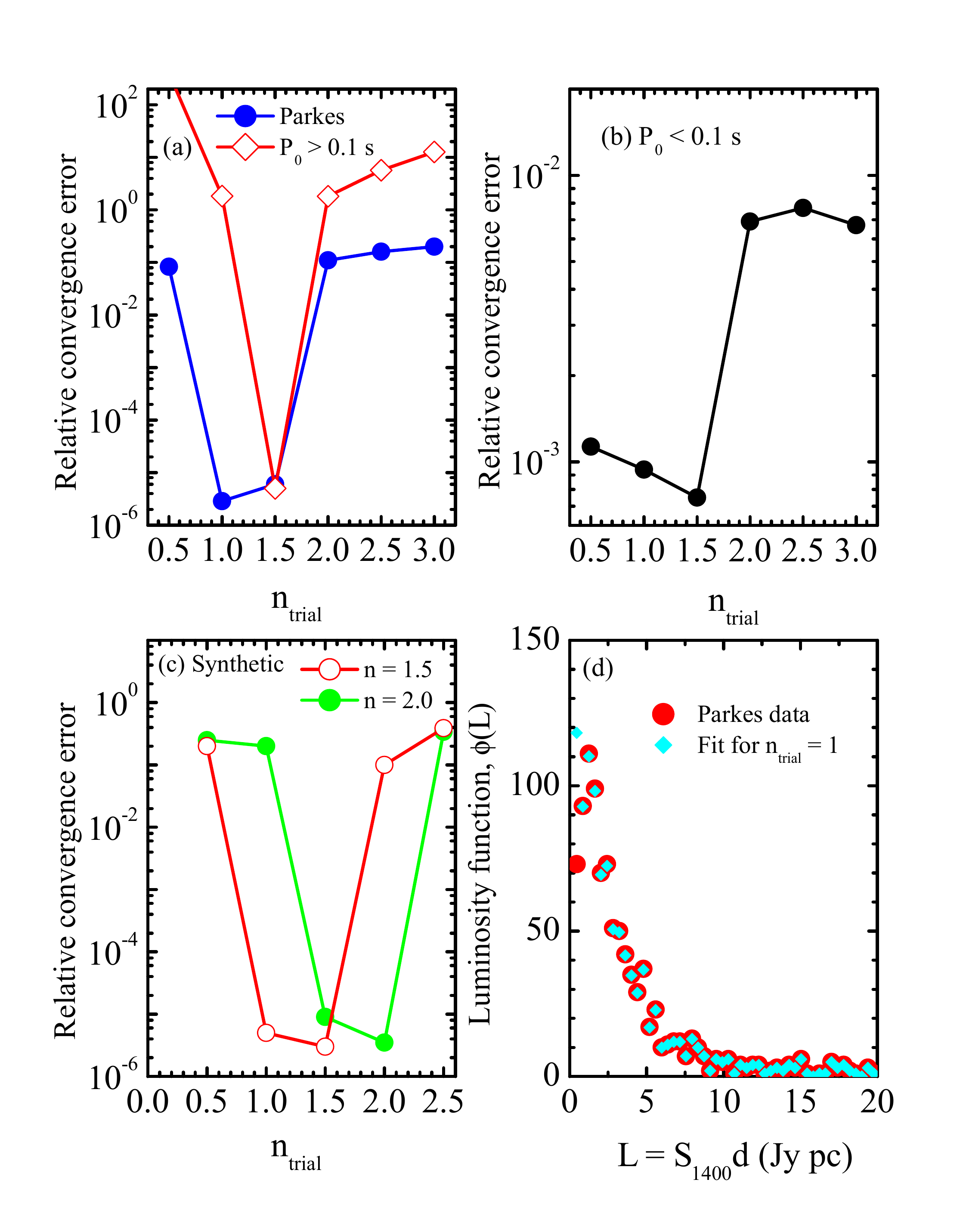}
\vspace{-8mm}
\caption{(a)~Blue solid points: relative convergence error
$\epsilon$ from the MLM
applied to 983 ($n_{\rm trial}=0.5,1$) 
or 980 ($n_{\rm trial} =1.5, 2, 2.5, 3$) pulsars from the 
Parkes Survey versus
trial exponent $n_{\rm trial}$.
Red, hollow diamonds: $\epsilon$ from the MLM
applied to 941 ($n_{\rm trial}=0.5,1$) 
or 938 ($n_{\rm trial} =1.5, 2, 2.5, 3$) pulsars
with periods $P_0>0.1$~s.
(b)~MLM fit for the 43 Parkes pulsars
with $P_0<0.1$~s. 
(c)~Relative convergence errors for synthetic galaxies in which
$S_{1400} \propto d^{-1.5}$ (red, hollow points) and 
$S_{1400}\propto d^{-2}$ (green, filled points)
versus $n_{\rm trial}$.
(d)~The inferred luminosity function for 983 pulsars with $n_{\rm trial} =1$; 
large circles are raw Parkes data; small diamonds are
the fitted luminosity function.}
\vspace{-5mm}
\label{astFig2}
\end{figure}

The MLM~\cite{george}
therefore finds that the observed 1400~MHz flux of pulsars
does not fall off as the conventionally-assumed $1/d^2$,
instead returning exponents of either 1 or 1.5.
There are two possible conclusions; either the dispersion
measure estimates of pulsar distance~\cite{dm3}
are radically (and consistently) incorrect by factors of order 10, or
pulsars do, in fact, have a flux that falls off more slowly with distance.
The former conclusion
would call into question the widely-used~\cite{giguere}
(but admittedly flawed) NE2001 method~\cite{dm3} for estimating the
distances of radio sources from dispersion measure, and also be at variance with
the presently-accepted distribution of matter within 
the Galactic plane~\cite{parkes,dm3}.
Instead, we suggest that $S_{1400}$ for pulsars
falls off more slowly with distance than the inverse-square law, 
due to the presence of a component whose flux varies as $1/d$.
This is in agreement with
the superluminal emission model of pulsars,
a concept that has also correctly predicted the emission
spectrum of the Crab and eight other pulsars~\cite{mnras,mnras2} 
and reproduced other
salient features ({\it e.g} apparent brightness temperatures,
Stokes parameters, small extent of the emitting region {\it etc.}) 
seen in pulsar observations~\cite{schmidt}.

This work is supported by the U.S. Department of Energy
(Los Alamos Laboratory-Directed Research and Development 
(LDRD 20080085DR) funds). 
We thank Joseph Fasel, Andrea Schmidt, Petr Volegov and 
William Junor for insights and encouragement.
We are grateful to Dr Elizabeth Kelly for guidance on the
statistics of truncated distributions.

\end{document}